# Deep Photonic Reservoir Computing with On-chip Nonlinearity

Jinlong Xiang, Youlve Chen, Yuchen Yin, Zhenyu Zhao, Chaojun Xu, An He, Xintong Lv, Yikai Su, and Xuhan Guo[*]

State Key Laboratory of Photonics and Communications, School of Information and Electronic Engineering, Shanghai Jiao Tong University; Shanghai, 200240, China.

*Corresponding author. Email: guoxuhan@sjtu.edu.cn

## Abstract

Reservoir computing, renowned for its low training cost, has emerged as a promising lightweight paradigm for efficient spatiotemporal processing. Increasing the reservoir depth brings enhanced learning capabilities that match top artificial intelligence models, but it remains challenging to realize deep photonic reservoir computing (DPRC) systems, due to the lack of scalable on-chip nonlinearity. Here, we introduce a versatile time-delayed DPRC framework that natively supports deep and concurrent spatiotemporal processing entirely in the optical domain. At its core, the system leverages free carrier dynamics in silicon microring resonators to provide the fundamental nonlinearity and short-term memory, and these nonlinear nodes are interconnected through true-time-delay lines that establish shared long-term memory. Benefiting from intrinsic physical nonlinearity and multi-timescale fading memory, this simple yet effective architecture demonstrates remarkable high-dimensional representation capabilities. On the NTU-RGB+D benchmark, the parameter-efficient DPRC system achieves superior action recognition accuracies compared to mainstream deep learning models, while requiring only a single-shot regression training procedure. We further verify a prototype DPRC chip that excels across diverse dataset classification and time-series prediction tasks. It enables a streamlined all-optical pipeline between hierarchical layers, delivering a consistent computational density of 334.25 TOPs/mm², independent of the reservoir depth and three orders of magnitude higher than conventional approaches. Moreover, its performance scales with near-zero hardware overhead by utilizing additional wavelength channels. This DPRC network is highly scalable on a silicon photonic platform, with flexible extension to hundreds of deep reservoir layers and parallel channels, paving the way toward intelligent optoelectronic systems for advanced real-time processing and parallel decision-making.

# Introduction

Artificial intelligence (AI) has demonstrated transformative capabilities, with recent breakthroughs in large language models even surpassing human-level performance in professional examinations[1]. However, this progress is shadowed by the soaring computational demand of ever-larger models, necessitating an urgent need for new computing alternatives. Optical computing has drawn increasing attention with exceptional advantages of broad bandwidth, low latency, and massive parallelism[2-4]. Notably, state-of-the-art photonic systems have already reduced processing times from the millisecond realm of biological neural systems to the picosecond scale, while achieving energy efficiency several orders of magnitude higher than leading electronic counterparts[5-8].

Among various neural networks, photonic reservoir computing (PRC) stands out as a compelling neuromorphic architecture for time-dependent information processing with significantly reduced hardware complexity[9]. PRC only requires training the readout layer through linear regression, thus circumventing the challenging gradient-based optimization needed in conventional photonic neural networks[10,11]. This training process is further facilitated by a single-shot output state acquisition that can be completed in minutes using photonic accelerators. Moreover, the core reservoir layer can be physically implemented with intrinsic dynamics in optical devices for nonlinear mappings, which can be more effective than mathematically-defined activation functions, such as Sigmoid. Following this route, the past few years have witnessed notable advancement in the field of PRC, with numerous spatially-distributed and time-delayed reservoirs demonstrated across free-space[12-15], fiber[16-22], and integrated photonic platforms[23-26]. By integrating the deep learning principle into deep PRC (DPRC) systems, both the linear and nonlinear memory capacity can be improved[27], resulting in substantially enriched representation diversity. It has even been theoretically proved that hierarchically coupled delay reservoirs can function equivalently to deep convolutional neural networks (CNNs)[28].

However, realizing scalable DPRC systems still remains a critical challenge, primarily due to the lack of integrated and efficient on-chip nonlinearity. To construct hierarchical layers, one straightforward approach is to reuse the same optical hardware to compute reservoir states layer-by-layer[20]. While easy to implement, this method introduces redundant latency and power overhead from repeated optical-electrical-optical (O-E-O) conversions. Another alternative is to build fully analog DPRC systems that bypass all digital processing and memory access. Representative implementations have been demonstrated using cascaded injection-locked semiconductor lasers[29] or exploiting phase modulation between two groups of frequency combs[30]. Nevertheless, these fiber-based setups are bulky, complex, and difficult to scale, limiting their practical deployment. From the perspective of long-term system cost and reliability, developing all-optical DPRC systems on integrated platforms appears most promising. However, existing on-chip PRC demonstrations, such as those based on linear coherent cavities[23], waveguide meshes[24,25], and the next-generation reservoir computing architecture[26], still largely depend on square-law photodetection to introduce nonlinearity, which fundamentally constrains the operating speed and scalability in DPRC architectures.

Here, we propose the first nonlinear time-delayed DPRC framework on a silicon photonic platform that allows hierarchical feature learning and concurrent multi-task processing in a streamlined all-optical data flow. Specifically, nonlinear free carrier dynamics, i.e., two-photon absorption and free-carrier dispersion, in microring resonators (MRRs) are leveraged to provide on-chip nonlinearity and short-term memory (Fig. 1b). Meanwhile, waveguide delay lines are incorporated to provide shared long-term memory across all nonlinear nodes. Benefiting from the physical nonlinearity and multi-timescale temporal processing, our DPRC network demonstrates exceptional spatiotemporal processing capability on skeleton-based human action recognition (Fig. 1a), one of the most actively studied challenges in computer vision. On the benchmark NTU-RGB+D dataset[31], an 8-layer DPRC system, with only ~1.1M learnable parameters, achieves leading accuracies of 98.1% and 96.7% in the standard cross-view and cross-subject evaluations, respectively. This performance outperforms mainstream deep learning models, such as graph convolutional networks (GCNs)[32-35] and transformers[36-38], while requiring only a single-shot regression training procedure (Fig. 1c). Experimental validation of our DPRC chip confirms excellent performance across diverse tasks of nonlinear dataset classification and time-series prediction. Entirely free from digital processing between successive layers, the prototype holds a consistent computational density of 334.25 TOPs/mm² independent of the reservoir depth, three orders of magnitude higher than existing PRC approaches (Fig. 5e). Moreover, this performance can be efficiently scaled with near-zero hardware overhead as additional wavelength channels all share the same long-term memory. Our highly scalable DPRC framework represents a promising candidate for next-generation neuromorphic hardware, opening up new avenues for ultrafast machine vision, robotic control, and high-speed communications.

## Results

**Principle of the DPRC framework**

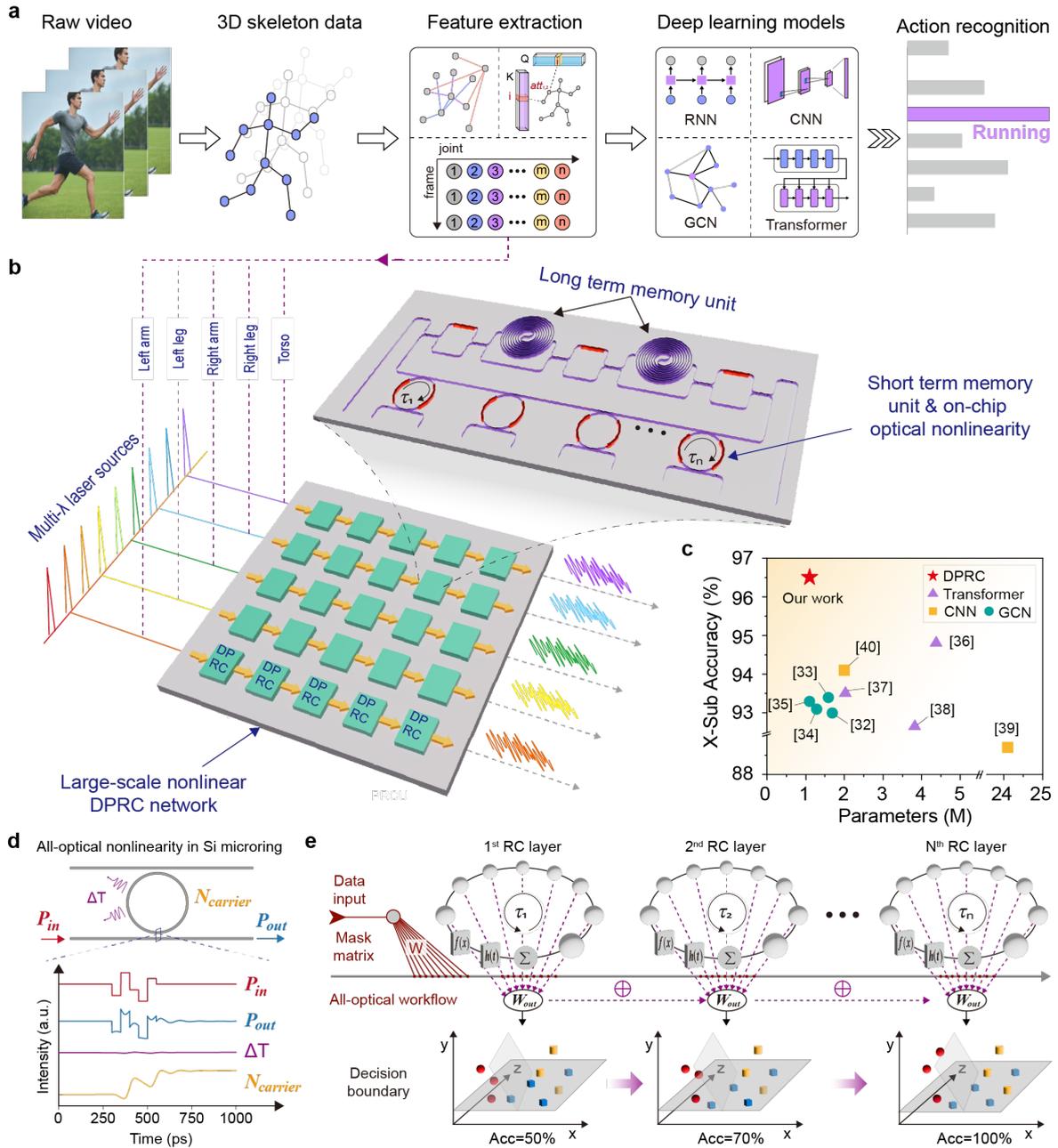

**Fig. 1 | Deep photonic reservoir computing (DPRC) framework with on-chip nonlinearity.**
**a**, Workflow of skeleton-based human action recognition using conventional deep learning models. **b**, Architecture of a large-scale nonlinear DPRC system, composed of clustered DPRC chips for streamlined deep and parallel processing entirely in the optical domain. Each DPRC chip incorporates spiral waveguide delay lines for long-term memory and microring resonators (MRRs) for nonlinearity and short-term memory. **c**, Performance comparison between an 8-layer DPRC system and state-of-the-art deep learning models on the NTU-RGB+D human action recognition benchmark. CNN, convolutional neural network. GCN, graph convolutional network. **d**, Nonlinear transient output response of a silicon MRR. Free-carrier effects provide short-term memory through decaying coupling between successive inputs, while thermal effect only induces a steady-state wavelength detuning. **f**, Working principle of a time-delayed DPRC system with enhanced high-dimensional representation capability.

As schematically shown in Fig. 1b, the large-scale nonlinear DRPC network comprises arrays of DPRC computing blocks monolithically integrated on a silicon photonic platform. Each row corresponds to a wavelength-multiplexed parallel channel, while each column represents a distinct reservoir layer. Preprocessed inputs, such as skeleton data from different body parts, are concurrently modulated onto different optical carriers and sequentially processed through hierarchically coupled DPRC layers. At its core, each DPRC chip implements a time-delayed reservoir design, where add-drop MRRs with dedicated input and output ports are interconnected via a closed-loop time delay line. Micro-heaters are embedded within MRRs to enable precise resonance tuning through localized thermal actuation, and the delay line is engineered with a compact, low-loss waveguide geometry (see Supplementary Note 1 for structure details). Furthermore, thermo-optical switches are integrated to enable programmable control over the overall delay time.

Under high-speed operation, the temporal response of an MRR reservoir is governed primarily by fast free-carrier dynamics, including carrier generation via two-photon absorption and resonance shifts induced by free-carrier dispersion, while slower thermal effects contribute only a quasi-static influence on the effective wavelength detuning. When the MRR works near its resonance with a relatively high pump power, abrupt changes in optical input power will lead to nonlinear modulation on the transmitted output signal (Fig. 1d). Meanwhile, the resulting carrier concentration requires a recovery time much longer than the data period to reach steady state. This slow relaxation process introduces temporal coupling between sequentially encoded virtual nodes, with exponentially decaying strengths, thus enabling the MRR to act as a nonlinear short-term memory element. To establish stable recurrent connections between computing nodes, delay lines are incorporated to provide long-term memory. Consequently, each DPRC chip exhibits rich multi-timescale fading memory, with characteristic times spanning from several picoseconds—determined by the photon lifetime within the MRRs—to several nanoseconds—dictated by both the physical length of delay lines and the carrier relaxation time in the MRRs, thereby significantly enhancing the representational diversity and nonlinear mapping capability.

Figure 1f conceptually illustrates the working principle of the DPRC system, where each reservoir layer utilizes a physical MRR to define multiple virtual neurons in a time-multiplexed manner. Input data are projected onto all virtual neurons via an input coupling matrix, and then undergo successive nonlinear transformations across each layer. Following a single pass of data transmission, the output states of neurons from all reservoir layers are collected to train a readout weight matrix. As visualized by the decision boundaries at the bottom, the transformed representations become increasingly linearly separable with greater reservoir depth, leading to improved classification accuracy. Notably, the entire process is accomplished through a streamlined end-to-end optical data flow, without any optoelectronic or digital-to-analogue conversion.

**Skeleton-based action recognition using DPRC**

Human action recognition plays a fundamental role in the computer vision community, with widespread practical applications in video surveillance, human-computer interaction, and robot vision. Among diverse data modalities, 3D skeleton data offers a compact and informative representation of the body structure with robustness against environmental changes, and can be easily collected with low-cost depth sensors. Despite decades of research using various deep learning models (Fig. 1a), such as CNNs[39,40], GCNs[32-35], and transformers[36-38], skeleton-based action recognition remains challenging due to the critical need for discriminative spatiotemporal feature extraction.

To highlight the exceptional processing capability of the DPRC system, we evaluate its performance on the NTU-RGB+D benchmark—a comprehensive and widely adopted dataset for skeleton-based action recognition[41]. The dataset encompasses 60 action categories, covering a diverse range of daily activities and interactions, e.g., drinking, jumping, and handshaking. The preprocessing pipeline is illustrated in Fig. 2a. Raw skeleton samples first undergo viewpoint-invariant normalization and are temporally aligned to 100 frames. To capture spatiotemporal dependencies effectively, the 25 body joints in each frame are grouped into five anatomical parts: torso, left arm, right arm, left leg, and right leg. Each part is then reshaped into a 2D feature map of size 3×D, with rows corresponding to spatial dimensions (x, y, z) and columns formed by concatenating coordinate vectors chronologically across frames. Finally, the feature map is converted into 1D vectors through multiplication with a randomly generated mask matrix $W^{D \times N}$, resulting in 15 input channels for the DPRC network.

As shown in Fig. 2b, preprocessed skeleton data are sequentially processed by $L_n$ DPRC layers distributed across 15 parallel computing branches. The outputs from all layers are collected and flattened into a state vector of dimension $L_n \times N \times 15$, where N denotes the number of virtual neurons per layer. The readout weights are trained using ridge regression with a regularization coefficient of $1 \times 10^{-5}$. Our evaluation follows the two standard protocols, Cross-Subject (X-Sub) and Cross-View (X-View), which assess generalization across different human subjects and camera viewpoints, respectively. Fig. 2c presents the results for a DPRC network with 150 neurons per layer. Recognition accuracy under both protocols increases consistently with the number of DPRC layers, hence confirming the enhanced representational capacity of deeper reservoir architectures. The per-category accuracy distributions for an 8-layer DPRC system are shown in Fig. 2d and 2e, achieving overall accuracies of 98.1% (X-View) and 96.7%

(X-Sub). Besides, the last few interactive action categories show lower accuracy than the average level, largely because these scenarios are modeled using single-subject skeleton data, which provides an incomplete representation of the interpersonal dynamics.

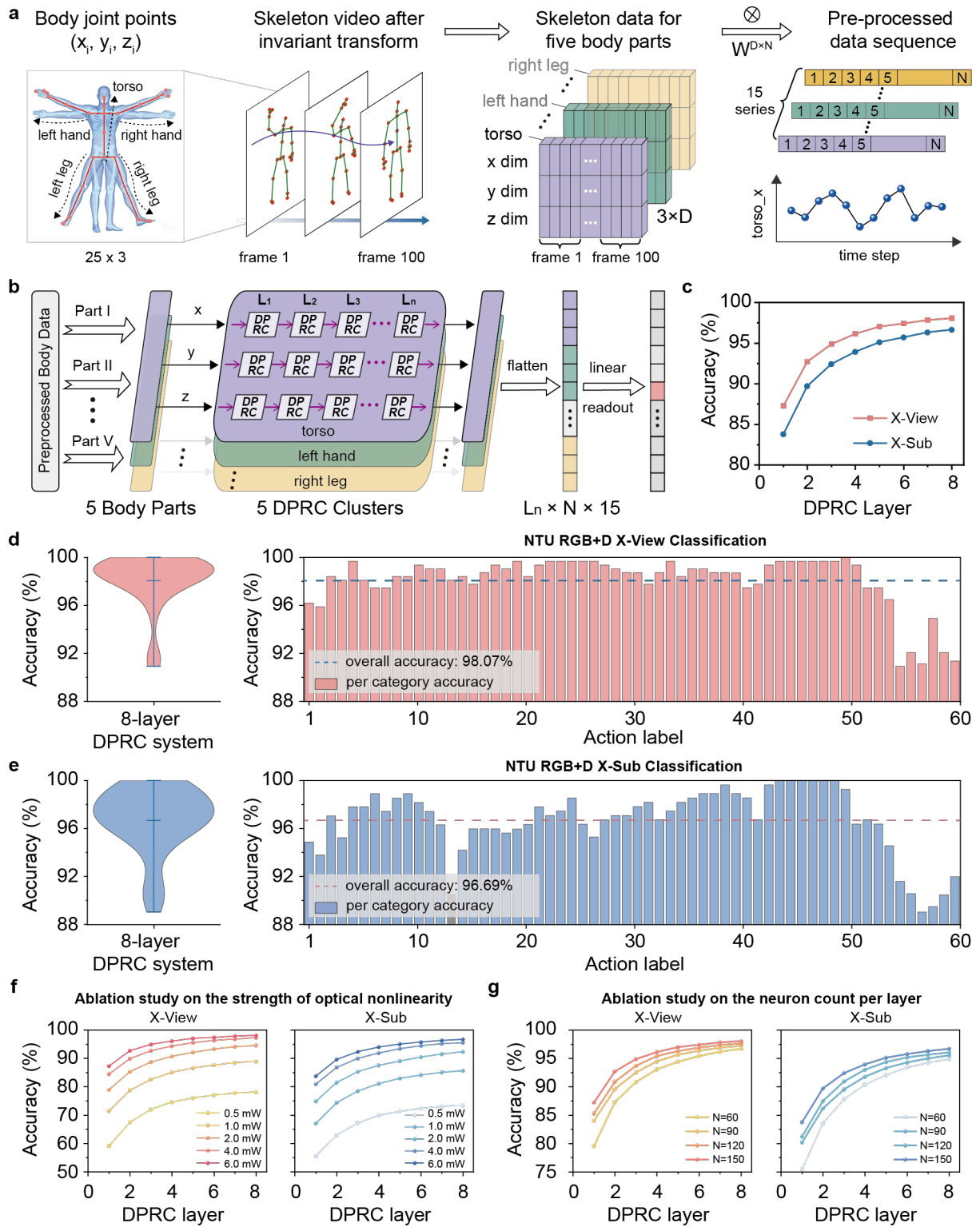

**Fig. 5 | Performance of DPRC systems on the NTU-RGB+D human action recognition benchmark. a,** Preprocessing pipeline and generation of input features from raw skeleton data. **b**, Framework of the DPRC system for skeleton-based action recognition. **c**, Dependence of

recognition accuracy on the number of DPRC layers, using 150 virtual neurons per layer. **d-e**, Per-category accuracy distributions for an 8-layer system, achieving overall accuracies of 98.1% (**d**, X-View) and 96.7% (**e**, X-Sub). **f**, Ablation study on on-chip nonlinearity strength, where the optical pump power is increased to induce stronger nonlinear effects in MRR reservoirs. **g**, Contribution analysis of the number of virtual neurons per layer on the recognition performance.

To investigate the role of on-chip optical nonlinearity, we evaluated the DPRC system under varying optical pump powers while keeping all other conditions constant. As shown in Fig. 2f, an 8-layer DPRC system operating at 0.5 mW performs even worse than a single-layer system driven at 6 mW. This result aligns with the principle that free-carrier effects scale with optical intensity. As the pump power decreases, the induced nonlinearity weakens substantially, leading to significantly degraded recognition accuracy. It should be noted that when the pump power reaches 8 mW, the MRR enters a self-pulsation regime in which thermal dynamics strongly influence the temporal response. The impact of neuron count per layer is further examined in Fig. 2g. The recognition performance of DPRC systems improves consistently with increasing virtual neurons, and gains little improvement beyond approximately 150 neurons, particularly for deeper architectures.

We systematically compared the DPRC framework with state-of-the-art deep learning models in Fig. 1c and Supplementary Table 1. By harnessing the intrinsic physical nonlinearity of MRRs, DPRC systems can extract spatiotemporal features more effectively than conventional models that rely on activation functions such as Sigmoid. Notably, an 8-layer DPRC system with only ~1.1M trainable parameters (150×8×15×60) achieves competitive top-1 accuracy on the NTU-RGB+D benchmark, outperforming mainstream GCNs[32-35] and transformers[36-38]. Moreover, training conventional models on digital hardware typically requires hundreds of iterations of gradient-based optimization, a process that grows increasingly time-consuming and energy-intensive with dataset scale. In contrast, the DPRC network employs a greatly simplified regression training procedure. The primary computational cost lies in the one-time acquisition of reservoir output states, which can be efficiently executed in minutes with underlying photonic hardware.

**DPRC prototype characterization**

To experimentally validate the potential of DPRC systems, we fabricated a DPRC prototype in a commercial silicon photonic foundry. Figure 3a shows the photograph of the DPRC chip after packaging, which integrates 6 MRR channels, 2 waveguide delay line units, and 3 Mach-Zehnder interferometer (MZI) switches within a compact footprint of 1.78 mm$^2$. Figure 3c-3e illustrate the enlarged photographs of spiral waveguides, MZIs, and a silicon MRR with an in-ring doped microheater, respectively.

To characterize the memory capacity of the DPRC chip, a rectangular pulse of 0.2 ns is injected into the first MRR, and the resulting output is recorded at the corresponding through port. As shown in Fig. 3f, there exist two distinct output pulses in the detected waveform, indicating the fading memory effect. The corresponding time delay, determined by the time difference between two power peaks, is approximately 1.45 ns. To further evaluate the nonlinear transient dynamics, a reference sequence of 20 Gbps is introduced into the same MRR pumped with an on-chip optical power of 6 dBm and a wavelength detuning of 50 pm. The measured waveform at output port O1 reveals a clear nonlinear modulation effect (Fig. 3g), thereby providing the essential on-chip nonlinearity required for DPRC systems.

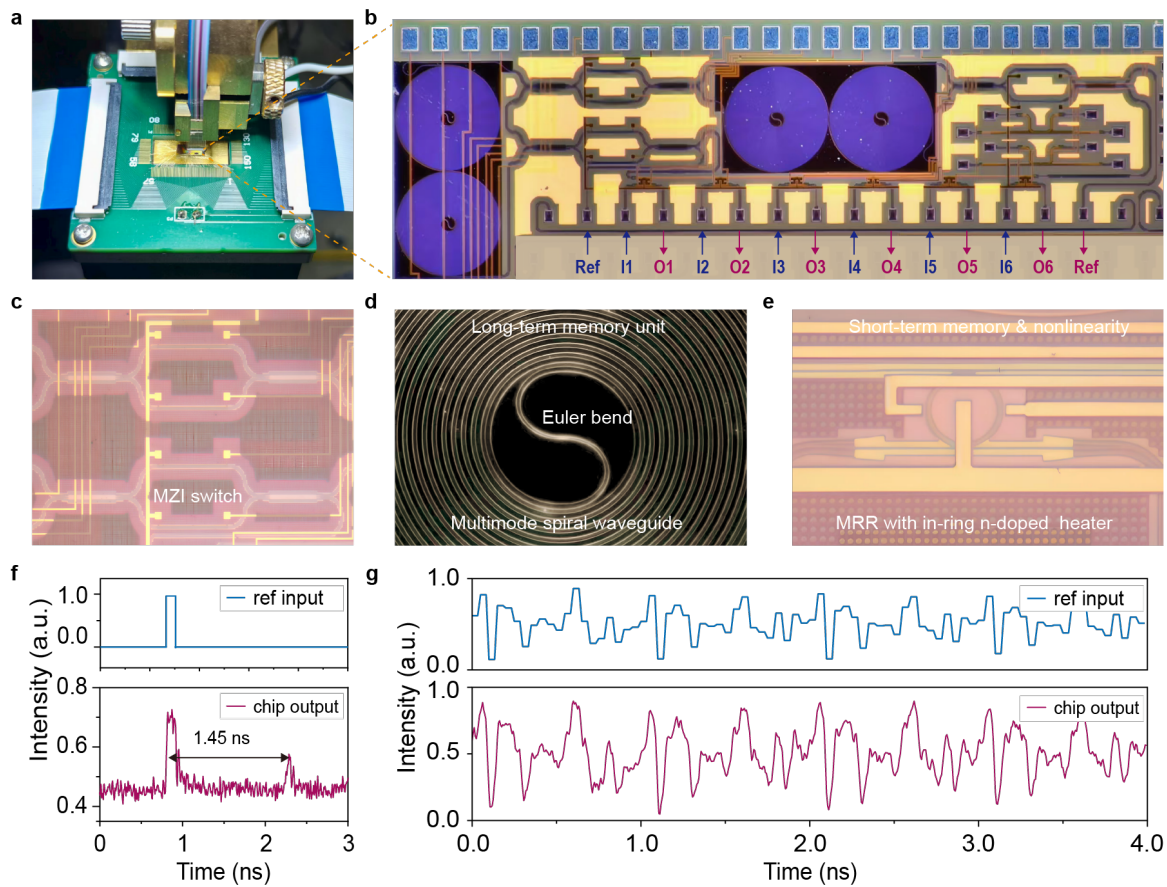

**Fig. 2 | Characterization of the DPRC chip. a,** Packaged DPRC chip with wire bonding. **b,** Microscope photonic of the fabricated chip, consisting of 6 MRR reservoirs, 2 delay units, and 3 MZI switches within a compact footprint of 1.78 mm$^2$. **c-e,** Zoom-in micrographs of two MZI switches, a spiral waveguide delay line, and a silicon MRR with an in-ring n-doped microheater. **f,** Characterization of the memory capacity by injecting a 0.2-ns rectangular pulse into the port I1 and recording the output waveform at port O1. A memory duration of 1.45 ns is indicated by the time difference between the two power peaks. **g,** Nonlinear temporal response of the first MRR when pumped with an on-chip optical power of 6 dBm and a wavelength detuning of 50 pm.

**Benchmark evaluations with shallow PRC**

We first experimentally evaluated the DPRC chip using shallow reservoir architectures on two benchmarks: Iris flower classification and Lorenz time-series prediction. The classic Iris dataset comprises 3 classes of 50 instances for each iris plant, with two species linearly inseparable from each other. The information processing flow is illustrated in Fig. 4a. Input features are encoded into a temporal sequence via a mask matrix and are then nonlinearly projected into a high-dimensional space by the DPRC chip. The resulting waveform, captured by an oscilloscope, is down-sampled to extract the output states for all samples. Following a 75%/25% training-test split, the final classification is performed through a winner-take-all readout. The influence of neuron number on classification accuracy is shown in Fig. 4b. Both training and test accuracies improve as the neuron count increases from 10 to 20, beyond which the test performance drops dramatically—a trend attributed to overfitting on the training samples. Confusion matrices for the optimal configuration of 20 neurons are presented in Fig. 4c and 4d. The accuracy reaches 97.3% and 97.4% on the training and test sets, respectively, thereby verifying the pattern recognition capability of the DPRC chip.

The MRR-based DPRC chip inherently supports wavelength multiplexing, which enables concurrent processing of distinct input features across multiple independent tasks. To validate this capability, the DPRC chip is configured to simultaneously predict all three components of a chaotic Lorenz attractor—a benchmark system known for its high sensitivity to prediction errors. As illustrated in Fig. 4e, the complex 3D trajectory is deconstructed into its constituent x, y, and z temporal components. Each signal is optically modulated onto a dedicated wavelength channel and processed independently within separate MRR reservoirs. After nonlinear transformations in the DPRC chip, the resulting reservoir states for each axis are delivered to a shared readout layer, which simultaneously generates one-step-ahead predictions for all three components. The prediction accuracy is quantified using the normalized mean square error (NMSE) between target and predicted values, defined as $NMSE = \frac{1}{L}\sum_{j=1}^{j=L}(Y_{tar}^j - Y_{pred}^j)^2 / var(Y)$. As shown in Fig. 4f and 4g, the predicted trajectories maintain excellent agreement with the ground truth during both training and testing phases, with negligible deviation. This is further confirmed by the low NMSE values of 0.006, 0.008, and 0.018 for the x, y, and z components, respectively. Moreover, this parallel computing architecture can be leveraged to improve the overall performance on a single task by aggregating outputs from multiple reservoir channels, as demonstrated on vowel recognition in Supplementary Note 5.

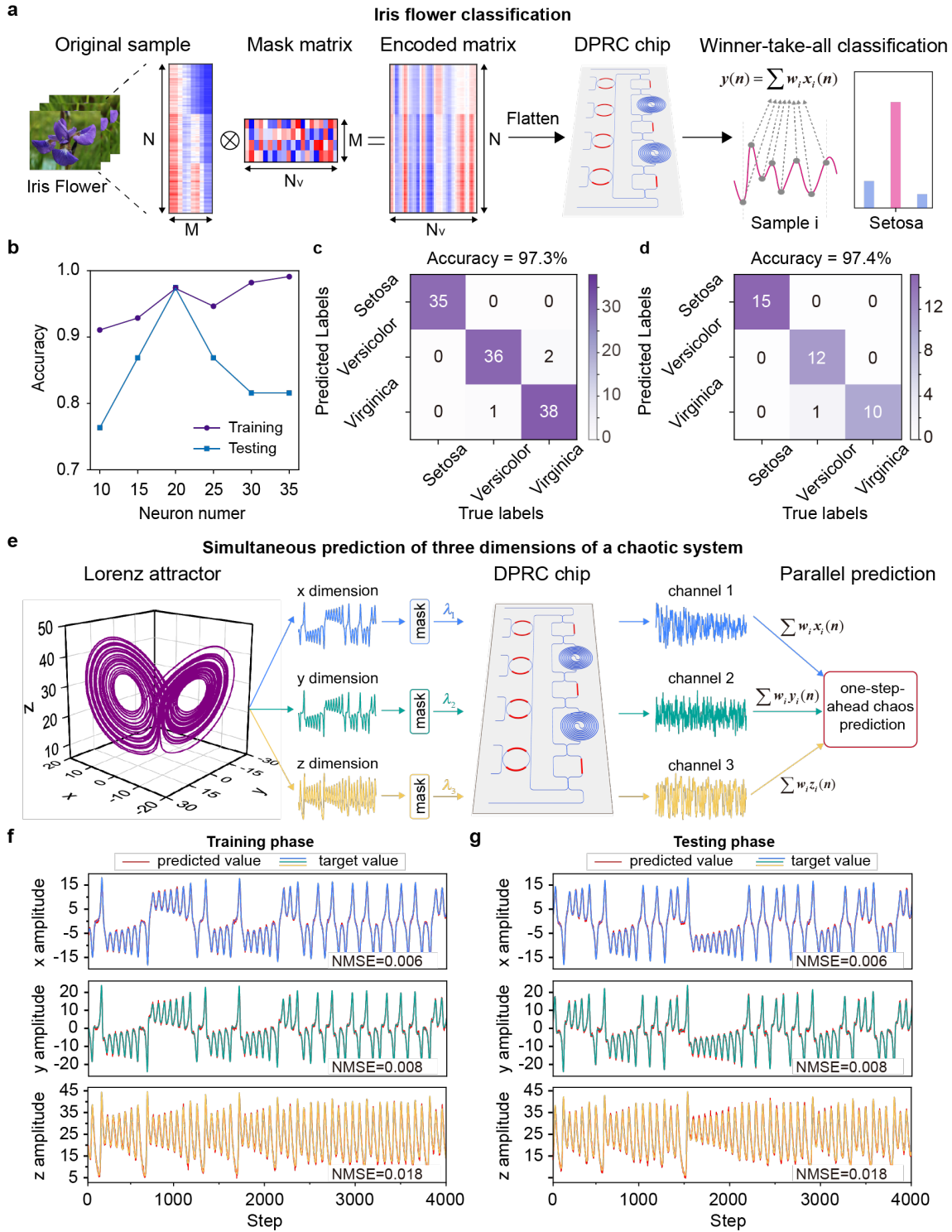

**Fig. 4 | Experimental validation on dataset classification and time-series prediction tasks with shallow PRC architectures. a–d**, Iris flower classification task. **a**, Processing workflow with a winner-take-all readout strategy. **b**, Classification accuracy as a function of the number of virtual neurons. **c, d**, Confusion matrices for the training **c** and **d** test sets using 20 virtual neurons. **e-g**, Parallel prediction of a chaotic Lorenz system. **e,** Workflow for the simultaneous prediction of all three components (x, y, z), where each component is processed by a dedicated MRR channel via wavelength-division-multiplexing. **f, g,** Comparison between the actual and predicted behavior during the **f,** training and **g,** testing phases.

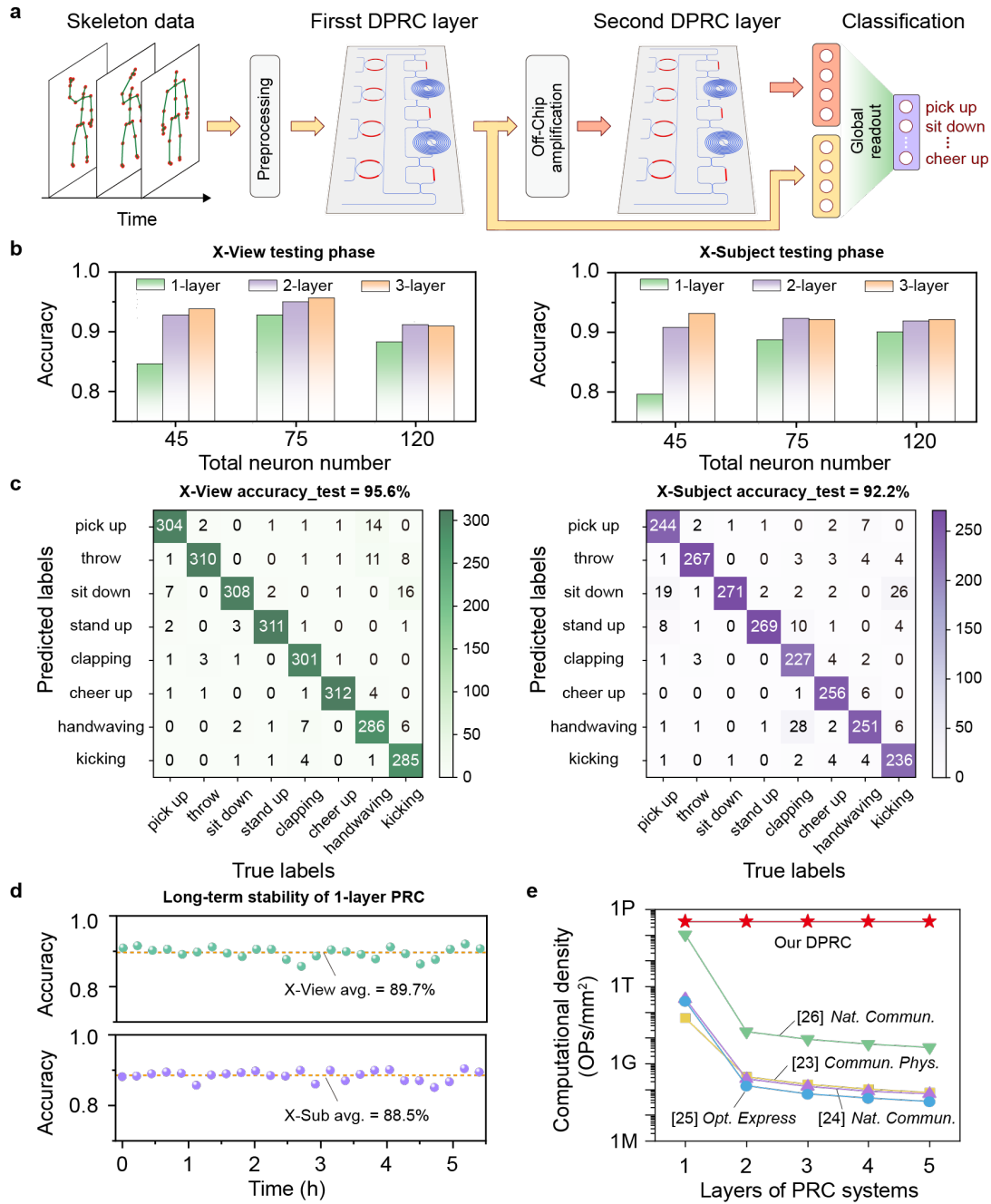

**Fig. 5 | Experimental demonstration of action recognition with DPRC. a**, Deep reservoir implementations using the DPRC prototype. The light is coupled off-chip to record the neuron states of each intermediate reservoir layer. **b**, Experiments with different reservoir depth and number of virtual neurons per layer. For deep hierarchical structures, pronounced performance gain is achieved in the case of fewer neurons, while diminishing for higher neuron counts. **c,** Confusion matrices for a 3-layer PRC system with 75 neurons per layer, achieving recognition accuracies of 95.6 % and 92.2% under the X-View and X-Sub evaluations, respectively. **d**, Long-term stability assessment over a 5.5-hour continuous measurement. **e**, Comparison of computational density across integrated PRC systems. The speed of existing approaches that rely on photodetection to introduce nonlinearity is fundamentally constrained by latency from

electronic readout and conversion, whereas the present DPRC supports seamless hierarchical processing through a fully optical data flow.

**Action recognition evaluations with DPRC**

To further demonstrate the performance enhancement by a deep hierarchical architecture, we implemented the NTU-RGB+D action recognition task on the DPRC prototype. Eight representative actions, including pick up, throw, sit down, stand up, clapping, cheer up, hand waving, and kicking something, are selected from the original dataset, resulting in 7,534 samples evaluated under the standard X-View and X-Sub protocols. Following the same pipeline in numerical simulations, the skeleton data are transformed into 15 feature vectors and concatenated into a single input sequence. As illustrated in Fig. 5a, the preprocessed data are modulated onto an optical carrier and sequentially processed through multiple cascaded MRR channels. The output of each intermediate reservoir layer is coupled off-chip to record the virtual neuron states, then amplified and fed back to drive the subsequent on-chip reservoir layer, with all MRR resonances aligned via thermal tuning. The recognition performance was assessed by training readout weights on neuron states collected from an increasing number of reservoir layers, enabling direct comparison of hierarchical depth on task accuracy.

As shown in Fig. 5b and 5c, the recognition accuracy under both protocols gets improved substantially with increasing reservoir layers. This performance gain is most pronounced with fewer neurons per layer (e.g., 45 neurons), and gradually diminishes at higher neuron counts (e.g., 120 neurons). Figure 5d and 5e present the corresponding confusion matrices of a 3-layer DPRC system with 75 neurons per layer, achieving accuracies of 95.6% and 92.2% under the X-View and X-Sub evaluations, respectively. Furthermore, we assessed the long-term system stability by repeating measurements over 5.5 hours using a single reservoir layer with 75 virtual neurons. The system maintained consistent performance (Fig. 5f) with average accuracies of 89.7% (X-View) and 88.5% (X-Sub), confirming its robustness for sustained operation. It should be noted that the off-chip amplification used in this proof-of-concept demonstration is not a fundamental need, and a fully integrated DPRC system could be realized by directly cascading MRR reservoirs on-chip.

## Discussion and conclusion

Increasing reservoir depth can substantially enrich the representation diversity and offer a compelling photonic analogue to deep CNNs. By eliminating the layer-wise O-E-O conversions, our DPRC architecture enables streamlined hierarchical processing entirely in the optical domain. As a result, the DPRC chip maintains a consistent computational density of 334.25 TOPs/mm² regardless of reservoir depth, as illustrated in Fig. 5e, three orders of magnitude higher than conventional PRC systems. Furthermore, this performance can scale seamlessly with minimal hardware overhead by integrating additional MRR channels that share the same long-term memory unit.

The maturity of silicon photonics, combined with advanced optoelectronic packaging, now enables the monolithic co-integration of all essential components for a fully integrated DPRC system. On-chip silicon nitride frequency combs can serve as high-quality multi-wavelength sources[42], while low-loss delay lines with high delay-density offer a path to substantially expand the memory capacity without compromising area efficiency[43]. Furthermore, the integration of semiconductor optical amplifiers[44] effectively mitigates propagation loss in deep PRC architectures, leading to significantly enhanced learning capacity of complex hierarchical features. Crucially, the remaining digital bottlenecks—the input mask layer and the linear readout layer—can be replaced by integrated photonic circuits, such as MZI meshes[45] and MRR weight banks[46]. The realization of such a standalone, all-analog PRC system would finally liberate photonic computing from the speed and power constraints of digital interfaces, fully unlocking its potential for transformative, real-world applications.

In conclusion, we report an integrated lightweight DPRC framework that enables all-optical hierarchical spatiotemporal processing and concurrent multi-task processing. Scalable on-chip nonlinearity is achieved by harnessing the intrinsic free-carrier dynamics in MRRs, eliminating the need for any optoelectronic conversion or the heterogeneous integration of exotic materials. The reservoir depth and parallel channels can be flexibly scaled on a silicon integrated platform, paving the way for wafer-scale intelligent photonic processors. The DPRC architecture delivers exceptional learning capability with extreme parameter efficiency, achieving top-tier recognition accuracies on the NTU-RGB+D benchmark. Compared to mainstream GCNs and transformers, the lightweight DPRC model with only ~1.1M parameters features dramatically simplified training, requiring only single-pass data acquisition and linear regression optimization. Furthermore, the fabricated DPRC prototype demonstrates excellent performance across various tasks in an all-optical pipeline, compared to existing approaches, offering three orders of magnitude improvement in processing speed within deep reservoir architectures. This work will not only inspire novel physical computing paradigms that harness intrinsic device dynamics for intelligent processing, but also pave the way for ultrafast photonic applications in robotics, communications, and sensing.

## Methods

### Chip fabrication

The PRC chip was fabricated on a silicon-on-insulator wafer with a 220 nm-thick silicon device layer, a 2 μm-thick buried oxide layer, and a 2.2 μm-thick oxide cladding. The MZI switches and MRRs were designed for single-mode operation, while the waveguide delay lines employed an optimized multimode design to reduce transmission losses.

**Data processing**

In the time-delayed DPRC architecture, an input mask is employed to distribute input samples across all virtual nodes. The mask elements are randomly drawn from the set [±0.3, ±0.6, ±1]. The resulting masked input signal is first normalized to the range of [-1, 1]. To compensate for the nonlinear sine transfer function of the Mach-Zehnder intensity modulator, the driving voltage is pre-distorted according to $V = \frac{V_\pi}{\pi} sin^{-1}(m_i u_i)$, spanning $[-\frac{V_\pi}{2}, \frac{V_\pi}{2}]$. To ensure precise time synchronization, a padding sequence containing three rectangular pulses is appended to the beginning of the encoded signal. After being processed by the DPRC chip, the output waveform is recorded and post-processed to train the readout weights via ridge regression.

**NTU-RGB+D benchmark**

The NTU-RGB+D dataset is a widely used benchmark for skeleton-based action recognition. It comprises 56,880 skeletal action sequences performed by 40 subjects across 60 action classes. The dataset provides two standard evaluation protocols: (1) Cross-Subject (X-Sub): Sequences are split based on subject IDs. Training data includes 20 subjects (IDs: 1, 2, 4, 5, 8, 9, 13, 14, 15, 16, 17, 18, 19, 25, 27, 28, 31, 34, 35, 38), while the remaining 20 subjects form the test set. (2) Cross-View (X-View): Samples are partitioned by camera view. All sequences captured by cameras 2 and 3 are used for training, while all sequences from camera 1 are reserved for testing.

**Time series prediction**

The Lorenz attractor is a classic chaotic system described by three coupled nonlinear ordinary differential equations:

$$\frac{dx}{dt} = \alpha(y - x)$$
$$\frac{dy}{dt} = x(\beta - z) - y$$
$$\frac{dz}{dt} = xy - yz$$

The model can describe diverse chaotic behaviors and physical phenomena depending on the parameters α, β, and γ. We generated chaotic trajectories using the standard parameter set (α = 10, β = 28, γ = 8/3) and initial conditions ($x_0$ = -5, $y_0$ = 0.2, $z_0$ = 20). The resulting 3D Lorenz attractor was then decomposed into its x, y, and z temporal components, with each component processed by a dedicated MRR reservoir.

**Experiment setup**

For the implementation of a single-layer PRC system (see Supplementary Fig. 3), the host computer pre-processes the input dataset (Iris flower dataset and Mackey–Glass time series) and sends the encoded sequence to a high-speed arbitrary waveform generator (AWG, Keysight M8195A, 25 GHz), which modulates the mask signal onto the optical carrier from a tunable continuous-wave laser (Santec, TSL770) via a Mach–Zehnder intensity modulator (Fujitsu, FTM7939EK, 25 GHz). Grating couplers are used to interface the bus waveguide of the MRR reservoir and single-mode fibers, with an optimized coupling loss of 5 dB/facet. To compensate for optical losses, two erbium-doped fiber amplifiers (EDFAs) are utilized to boost the optical power at the input and output of the PRC chip. A tunable filter centered at the pump wavelength with a passband of 0.8 nm is added behind the EDFA to improve the signal-to-noise ratio. After that, the filtered light is sequentially captured by a commercial photodetector (Finisar, XPDV21x0, 50 GHz) and digitized by a real-time oscilloscope (Lecroy, LabMaster 10 Zi-A, 65 GHz, 160 GS/s). Finally, the acquired waveform is sent back to the computer and used to train the linear readout layer. In all experiments, the time interval per virtual node is fixed at 50 ps, corresponding to a data rate of 20 Gbps. The optimal number of virtual neurons per layer is task-dependent, as the expected nonlinearity and memory capacity vary across applications.

For the parallel prediction of the Lorenz attractor, its three components are modulated onto separate wavelengths ($\lambda_1$ = 1549.58 nm, $\lambda_2$ = 1548.60 nm, and $\lambda_3$ = 1546.72 nm), each processed by a dedicated MRR reservoir. The state acquisition and readout training are then performed independently for each channel.

For the implementation of a deep PRC system, we first apply thermal tuning to coarsely align the resonance wavelengths of MRRs. The output from an intermediate MRR is coupled off-chip, amplified by an EDFA, and split into two branches. A 10% tap is directed to a photodetector and digitized by the oscilloscope, providing the virtual neuron states for the corresponding reservoir layer. The remaining 90% of the optical power is fed back—after polarization control—into the DPRC chip to drive the subsequent MRR, thereby closing the loop for deeper hierarchical processing.

## Acknowledgments

This work was supported by the National Research and Development Program of China (2023YFB2804702); National Natural Science Foundation of China (NSFC) (62175151, 62341508，62405185，62505175 and 625B2113); Shanghai Science and Technology Innovation Action Plan (25LN3201000 and 25JD1405500); Shanghai Municipal Science and Technology Major Project.

## Author contributions:

X.H.G. initiated and supervised the project. X.H.G. and J.L.X. conceived the research. J.L.X. designed and characterized the silicon chip. J.L.X., Y.L.C., Y.C.Y., and X.T.L. conducted the experiments. J.L.X. and Y.L.C. processed the data. All authors analyzed the results and contributed to the manuscript draft.

## Competing interests

Authors declare that they have no competing interests

## Data availability

The NTU-RGB+D, Iris, and digit image datasets are all publicly available and can be downloaded online from previous work. Source data are provided with this paper.

## Code availability

The software code is available